\documentclass[aps,prl,twocolumn,superscriptaddress,nobalancelastpage,showpacs]{revtex4}

\usepackage{graphicx}% Include figure files
\usepackage{dcolumn}% Align table columns on decimal point
\usepackage{bm}% bold math
\usepackage{longtable}
\begin{document}

\title{Field induced tricritical behavior in the $S=1/2$ quasi one-dimensional frustrated Ising antiferromagnet}

\author{E. Meloche and M. L. Plumer}
\affiliation{Department of Physics and Physical Oceanography, Memorial University of Newfoundland, St John's, Newfoundland, Canada A1B 3X7}

\date{\today}% It is always \today, today,
%\email{eric@physics.mun.ca}

\begin{abstract}
The results of extensive histogram cluster heat-bath Monte Carlo simulations on the critical behavior of the quasi-one dimensional Ising antiferromagnet on a stacked triangular lattice are presented.   A small applied field is shown to induce a crossover from XY universality to mean-field tricritical behavior. Experimental estimates of critical exponents suggest that these two types of phase transitions are observed in $S=1$ CsNiCl$_3$ and $S=1/2$ CsCoBr$_3$, respectively.  The present results demonstrate that this difference can be explained by an unusual staggered magnetic field arising from quantum exchange mixing previously proposed to account for spin excitations in $S=1/2$ quasi-one-dimensional Ising antiferromagnets. 
\end{abstract}

\maketitle

Meaningful comparisons between experimental and theoretical results for critical exponents can be notoriously difficult due to  inadequacy of standard techniques to sample thermodynamic quantities close enough to the phase transition region.  It is often the case that small perturbations such as weak anisotropy govern the universality of critical phenomena but only very close to the critical temperature ($T_N$), rendering access through simulations or measurements problematic.  Such issues are exacerbated by critical behavior thought to be influenced by magnetic frustration \cite{diep}.  Additional complications can arise in model systems where exchange interactions are strongly anisotropic or in cases where quantum effects govern spin dynamics.  The Ising antiferromagnet on a stacked triangular lattice (ISTAF) with strong c-axis exchange and weak inter-chain exchange interactions is an example of such a model system. Many quasi-one-dimensional (1D) ABX$_3$ compounds exhibit strong short-range intra-chain magnetic order (SRO) at temperatures well above the onset of true three-dimensional (3D) long range order (LRO) \cite{collins}.  The present work demonstrates that quantum effects which govern 1D spin dynamics can impact effective critical properties of the 3D system.  

Exchange mixing due to an unquenched orbital moment and crystal field effects in quasi-1D $S=1/2$ ISTAF's has been proposed to give rise to an unusual contribution to the effective Hamiltonian in the form of a staggered magnetic field with a periodicity of two along the c-axis. This term has been employed to explain low temperature neutron scattering spectra in CsCoCl$_3$ and CsCoBr$_3$ \cite{Nagler1} and Raman data in TlCoCl$_3$ \cite{kuroe}.  Invoking such an effect relies on the well-developed SRO intrinsic to quasi-1D magnetic systems at low T.  Its applicability has been questioned, however, due to the fact that the inclusion of such a term in the model Hamiltonian essentially assumes 1D LRO and is thus a mean-field approximation inappropriate for soliton dynamics \cite{goff}.  The impact of this field term on the measurements of {\it effective} critical behavior observed close to $T_N$ is not subject to this restriction and deserves consideration even though it may be strictly irrelevant in the renormalization group sense {\it at} $T_N$.  In this work, the effect of this quantum staggered field in the $S=1/2$ quasi-1D ISTAF is mimicked in Monte Carlo (MC) simulations by considering the influence of a small uniform magnetic field applied to a model with ferromagnetic intra-chain interactions with AF inter-chain coupling.  No such effective quantum staggered field is predicted to occur in the case of non-$S=1/2$ STAF's such as ($S=1$) CsNiCl$_3$ which has been shown to exhibit integer-spin Haldane-gap phenomena \cite{buyers}.  
 
   Critical behavior of the ISTAF have been investigated using numerous theoretical approaches, MC simulations and experimentally. Despite the large number of studies there is still no clear consensus in the literature regarding the critical exponents characterizing the phase transition between the partially disordered phase (PD) where one of the three sublattices remains disordered, and the paramagnetic phase, at $T_N$.  Most modeling results \cite{collins,Plumer1}  support the notion from symmetry arguments that the transition at $T_N$ belongs to the $XY$ universality class \cite{blankschtein}.  There have been suggestions, however, from several MC simulations that critical exponents are close to mean-field tricritcal \cite{Heinonen1,Koseki1}, but limited statistics and data analysis were used in these investigations.   From the experimental standpoint, a number of past neutron scattering studies of CsCoBr$_3$ and CsCoCl$_3$ \cite{collins} have obtained sets of critical exponents that largely support the idea of $XY$ critical behavior.  Intriguingly, recent high resolution neutron scattering experiments on CsCoBr$_3$ by Mao {\it et al.} \cite{Mao1} revealed results which suggest with tricritical mean-field behavior and it was speculated that this could be attributed to the anisotropic nature of the exchange interaction or a consequence of the quantum nature of the $S=1/2$ spins.  Experimental data on weakly axial $S=1$ CsNiCl$_3$ with a quenched orbital moment \cite{buyers} are consistent with $XY$ universality \cite{collins,guy2}.    Both types of experimental systems, strongly Ising and weakly Ising, exhibit magnetic transitions at $T_N$ with the same symmetry and thus should belong to the same universality class. 
    
   Motivated by the new experimental data of Mao {\it et al.}, extensive MC simulations using the efficient Cluster Heat Bath algorithm (CHB)\cite{Koseki2,Matsubara1}, combined with the histogram method\cite{Ferrenberg1}, were used here to investigate the effects of anisotropic exchange as well as a small applied field on the critical properties of the classical ISTAF.  This work serves to the extend previous MC simulations of the ISTAF mentioned above, but especially that of Koseki {\it et al.} \cite{Koseki1}, where the CHB was used with strongly anisotropic exchange but with no applied field, as well as that of Netz and Berker \cite{Netz1} and Plumer and Mailhot \cite{Plumer1} who considered \textit{isotropic} exchange interactions with a nonzero applied field.   The present results also compliment MC simulations on the frustrated XY-STAF where anisotropy in the exchange interactions was shown to induce a first-order transition \cite{Plumer2} in agreement with recent experimental data \cite{guy}.

MC simulations in the present work used the order-parameter culumant crossing method\cite{Peczak1} to determine the critical temperature of the system. Finite-size scaling analysis at $T_N$ was then used to extract estimates of the critical exponents $\mu, \gamma$ and $\nu$.  Both of these techniques have been thoroughly tested on frustrated Ising and Heisenberg spin models \cite{Peczak1}. 

% In zero field the results for the critical exponents obtained for 4 different models are consistent with 3D XY universality and in agreement with the theoretical predictions of \cite{blankschtein}.  The sensitivity  of the critical exponents with respect to the choice of transition temperature is examined.  Our results show that critical exponents characteristic of tricritical mean-field behavior can be obtained if the finite size scaling analysis is performed at a temperature slightly lower than the estimated $T_N$. 

For the purposes of determining the classical critical behavior it is adequate to use a model system with ferromagnetic exchange interactions along the $c$-axis, $J_0$, with or without a small applied uniform field $H$ directed along the $c$-axis. In addition to near-neighbor  AF interactions between chains, $J_1$, the effects of a small next-nearest-neighbor  exchange coupling, $J_2$, was also considered as in previous MC simulations \cite{Plumer3}.  These effects can be incorporated in the Ising Hamiltonian:
\begin{eqnarray}
\mathcal{H}= -\sum_{<i,j>} J_{ij} \sigma_{i} \sigma_{j}-H\sum_{i} \sigma_{i}
\end{eqnarray}
\noindent  where the spin at any site is $\sigma_{i}=\pm 1$.  In zero applied field 4 different cases were investigated with parameters listed in table \ref{cases4}.  For all cases $J_0$ and $J_2$ are ferromagnetic and $J_1$ is antiferromagnetic.  Finite-size scaling analysis was also performed at 5 nonzero values of $H$ for case II. 
\begin{table}[ht]
\vspace{-0.1cm}
\begin{center}
\caption{\label{cases4} Exchange parameters used for the 4  zero-field cases.}
\vspace{-0.2cm}
\begin{tabular}{ccccccc}
\\
\hline\hline\
Case& $J_0$ & $J_1$ & $J_2$ \\
\hline\hline\
I& 1.0& -1.0&0\\
II& 10.0& -1.0&0\\
III& 10.0& -1.0&0.1  \\
IV& 10.0& -0.3&0.001  \\
\hline\hline
\end{tabular}
\end{center}
\vspace{-0.55cm}
\end{table}
%\begin{figure}
%%\includegraphics[width=0.45\textwidth]{i1.eps}\includegraphics[width=0.45\textwidth]{OPcum10651.eps}\\
%\includegraphics[width=0.45\textwidth]{OPcum3.eps}\includegraphics[width=0.45\textwidth]{kosekiTc.eps}
%\caption{\label{figcumulant} Results of applying the order-paramater cumulant crossing method to estimate the critical temperature $T_N$.  Figs (a)-(d) corresponds to cases I-IV, respectively.  The straight lines correspond to linear fits to the data with $\textrm{ln}^{-1}(L'/L)\leq 2.2$.  } 
%\end{figure}
\begin{figure}[ht]
\vspace{-0.3cm}
\includegraphics[width=0.40\textwidth]{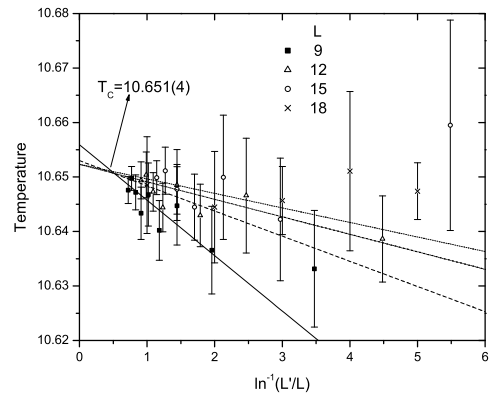}\\
\vspace{-0.5cm}
\caption{\label{figcumulant} Results of applying the order-paramater cumulant crossing method to estimate the critical temperature $T_N$ for case II.  The straight lines correspond to linear fits to the data with $\textrm{ln}^{-1}(L'/L)\leq 2.2$.  } 
\vspace{-0.6cm}
\end{figure}
The isotropic case I ($|J_0| = |J_1|$) was previously investigated using the MC histogram method with periodic boundary conditions for an $N=L\times L \times L$ system \cite{Bunker1,Plumer1,Plumer3}.    Different sets of critical exponents characteristic of different universality classes were obtained in these previous studies, serving to illustrate the sensitivity of analyzing MC simulation data for the frustrated Ising model.   Cases II, III, and IV correspond to systems with anisotropic quasi-1D exchange interactions (with $|J_0| \gg |J_1|$) for different values of $J_2$.   The anisotropic case IV was previously investigated\cite{Koseki2} using the CHB algorithm and estimates of the critical temperature and the critical exponents $\beta$ and $\nu$ were obtained using a data collapse method.   

The CHB algorithm employs open boundary conditions along the $c$-axis and periodic boundary conditions in the other directions. For systems with quasi-1D exchange interactions the CHB algorithm is more efficient than the Metropolis algorithm and allows for better statistics when simulating larger lattice sizes.   We considered anisotropic lattices with $N=L\times L \times 10 L$  and  $L=9,12,...,33,36$.   These system sizes are smaller than those employed in Ref.\cite{Koseki1}. However, a significantly larger number of MC steps were used in any particular run.   Averages were also performed over 10 independent simulations using different random initial spin configurations. Estimates of the errors were obtained by taking the standard deviation from the different simulations. Runs of $1\times 10^5-8\times 10^5$ MC steps were used for equilibration, and $5\times 10^5-1.2\times 10^6$ MC steps were used to calculate averages of several thermodynamic quantities including the  order parameter $O$, susceptibility $\chi_1$, specific heat $C$, energy cumulant $U_E$, order-parameter cumulant $U_M$ and first logarithmic derivative of the order parameter $V_1$, as defined in Ref. \cite{Peczak1}. The primary order-parameter $O$ is defined in terms of the $\mathbf{Q}_{th}$ Fourier component of the spin density as $O=\left|\sum_i \sigma_i \exp(i \mathbf{Q}\cdot\mathbf{R}_i)\right|/N$
%\begin{eqnarray}
%\chi_1&=&\frac{N}{T}\big(\langle O^2\rangle-\langle O\rangle^2\big)\nonumber\\
%\chi_2&=&\frac{N\langle O^2\rangle}{T}\nonumber\\
%%C&=&\frac{1}{NT^2}\big(\langle E^2\rangle-\langle E\rangle^2\big)\nonumber\\
%U_E&=&1-\frac{\langle E^4\rangle}{3\langle E^2\rangle^2}\nonumber\\
%U_M&=&1-\frac{\langle O^4\rangle}{3\langle O^2\rangle^2}\nonumber\\
%V_1&=&\frac{\langle OE\rangle}{\langle O\rangle}-\langle E\rangle
%\end{eqnarray} 
%\begin{eqnarray}\label{opeq}
%%O=\frac{1}{N}\left|\sum_i \sigma_i \exp(i \mathbf{Q}\cdot\mathbf{R}_i)\right|
%\end{eqnarray}
with $\mathbf{Q}=(2\pi/3,2\pi/3,0)$ in units of the lattice constants associated with the simple hexagonal structure. In addition, the temperature dependence of the secondary order-parameter $O'$ with the wavevector $\mathbf{Q}=(0,0,0)$, corresponding to the uniform magnetization, was also examined. Relevance of this component of the spin density (in zero applied field) on the nature of the critical behavior in these systems has been previously speculated \cite{Heinonen1,Koseki1}.  

Accurate estimation of the critical temperature is an essential first step in utilizing the histogram method for determining critical exponents. Temperature sweeps for the $L=24$ system were initially performed using fewer MC steps to obtain a rough estimate of the transition temperature $T_N$ by locating the maxima of the susceptibility $\chi_1$ and specific heat $C$.  For example, for case IV, temperature scans from $T=6.8-8.0$ in increments of $\Delta T=0.05$ yielded an estimate of $T_N=7.40$. To determine the critical temperature more accurately, several histograms were generated at temperatures above and below $T_N$.   In this case, histograms were generated at $T=$7.34, 7.36, 7.38, 7.40, 7.42 for each lattice size $L$.  
\begin{figure}
\begin{center}
\includegraphics[width=0.4\textwidth]{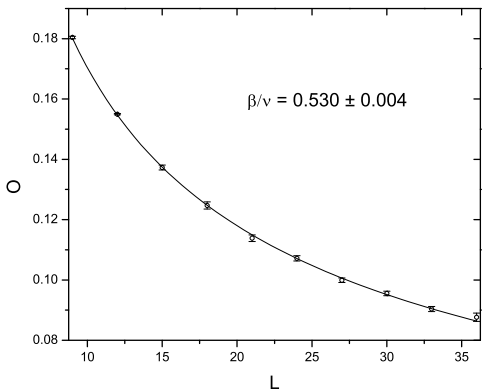}
\vspace{-0.5cm}
\caption{\label{fig6}
Finite-size scaling of the order parameter $O$ for case II where error bars on the data represent the standard deviation of the ten runs.} 
\end{center}
\vspace{-0.90cm}
\end{figure}
The order-parameter cumulant crossing method was used for the sets of parameters of Table I and an illustrative example of the results for case II is presented in Fig. \ref{figcumulant}.  For each case we plot the temperature at which $U_M$ for lattice size $L'$ intersects with the cumulants for $L=9,12$ and $15$.   Linear fits are made using the results in the asymptotic region (i.e., $\textrm{ln}^{-1}(L'/L)\leq 2.2$) and an estimate of the critical temperature $T_N$ is obtained from the average value of the crossing points.   The error $\pm \Delta T_N$ represents the standard deviation of these values. 

%Contrasting with the previous MC results\cite{Plumer1} the intersection points between $U_M$ at $L'$ and $L$ for the case with isotropic exchange (case I) occur over a much narrower temperature range due to the anisotropy of the lattice.  

%In Fig. \ref{figsop} we plot the temperature dependence of the primary and secondary order parameters $O$ and $O'$ for case II %parameters and lattices with $L=12$ and $L=24$.   We find that the secondary order-parameter $O'$ is irrelevant at the %transition temperature $T_N$ for all four cases investigated  

%\begin{figure}
%\begin{center}
%\includegraphics[width=0.45\textwidth]{sop.eps}\\
%\caption{\label{figsop} Temperature dependence of the (a) primary order-parameter $O$ and (b) secondary order-parameters $O'$ defined in Eq. (\ref{opeq}).  Results for case II are shown with lattice sizes $L=12$ and $L=24$.  } 
%\end{center}
%\end{figure}

%\begin{figure}
%\begin{center}
%\includegraphics[width=0.45\textwidth]{ecum12.eps}\\
%\caption{\label{fig3} Scaling of the energy cumulant $U_E$ as a function of lattice size $L$ is shown for the different cases.  The lines joining the data are guides to the eye and the horizontal line denotes $U_E=2/3$. } 
%\end{center}
%\end{figure}

The fourth-order energy cumulant evaluated at $T_N$ as a function of system size $L$ was also calculated in each of the four cases.   The results extrapolate to $U_E^*=0.666663(3)$ for $L\rightarrow \infty$ in all cases, as expected for a continuous phase transition. 
\begin{table}[ht]
\vspace{-0.1cm}
\begin{center}
\caption{\label{critexp} Comparison of the critical temperature and critical exponents determined in this work with other MC studies.}
\vspace{-0.2cm}
\begin{tabular}{ccccccccc}
\\
\hline\hline\
& &$T_N$ & &$\beta$ & &$\gamma$ && $\nu$ \\
\hline\hline\
I&& 2.926(3) && 0.344(7) && 1.31(3)&&0.671(9)\\
Bunker et al.\cite{Bunker1}&&  2.920(5) && 0.311(4)  && 1.43(3)&&0.685(3)         \\
Plumer et al.\cite{Plumer1}&&  2.9298(10) && 0.341(4)  && 1.31(3)&&0.662(9)         \\
Heinonen et al.\cite{Heinonen1}&&  2.88 && 0.19(1)  && 1.15(5)&&         \\
\hline
II&& 10.651(4)&& 0.355(6) && 1.33(3) && 0.670(7)\\
\hline
III&&11.998(4) &&0.358(5) && 1.28(3)&& 0.677(3)\\
\hline
IV&&7.406(6) && 0.362(7)&&1.35(2) && 0.673(4)\\
Koseki et al.\cite{Koseki2}&&  7.34(4) && 0.21(1)  && 1.31(3)&&0.70(3)         \\
\hline
XY&& &&0.345&&1.316&&0.671\\
Tricritical &&&&1/4&& 1 && 1/2\\
\hline\hline
\end{tabular}
\end{center}
\vspace{-0.8cm}
\end{table} 
Finite-size scaling analysis at the critical temperature $T_N$ of the various thermodynamic quantities $O \sim L^{-\beta/\nu}$, $\chi \sim L^{\gamma/\nu}$, and  $V_1 \sim L^{1/\nu}$, were used to obtain estimates of critical exponent ratios.   Representative results are shown in Fig. \ref{fig6} for the order parameter in case II where the error on the exponent ratio is obtained from the robustness of the fit.  Our results show that the secondary order parameter $O'$ is not relevant at $T_N$ for $H=0$.

In Table \ref{critexp} the zero field estimates of the critical temperature and critical exponents determined in this work are compared with those obtained by other MC studies.   For each case studied, our set of critical exponents indicate 3D XY universality.  This is in agreement with symmetry arguments as well as the MC results of Ref. \cite{Plumer1} but contrast with the tricritical behaviour seen in Refs. \cite{Heinonen1} and \cite{Koseki2}, where less intensive simulations were used.

\begin{figure}
\vspace{-0.4cm}
\begin{center}
\includegraphics[width=0.4\textwidth]{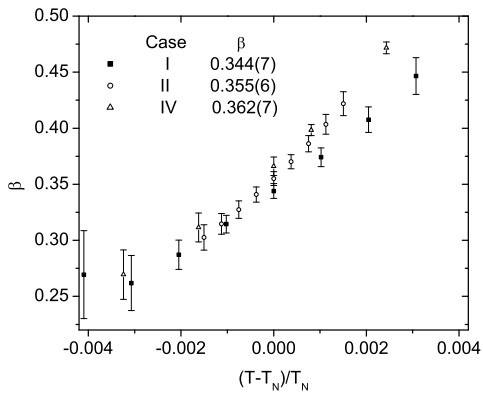}
\vspace{-0.3cm}
\caption{\label{fig2} Values of the critical exponents $\beta$ for cases I, II, and IV, obtained for different choices of the critical temperature. Results for case III are qualitatively similar and have been omitted for clarity.} 
\end{center}
\vspace{-0.85cm}
\end{figure}

In order to estimate errors due to the uncertainty in the critical temperature $T_N$, finite-size scaling analysis was performed at temperatures slightly above and below $T_N$.   In Fig. \ref{fig2}, results are shown for the critical exponents $\beta$, $\gamma$ and $\nu$ versus the choice of critical temperature for cases I, II, and IV.  We find that the cases with quasi-1D exchange interactions (cases II, and IV) are more sensitive to the choice of critical temperature than the case with isotropic exchange coupling (case I).   These results indicate that the critical temperature region is extremely narrow for the quasi-1D models, highlighting the need for care in the analysis and interpretation of both computational and experimental data for such systems.
\begin{table}[ht] 
\vspace{-0.1cm}
\begin{center}
\caption{\label{critexp2} Critical temperature and critical exponents for several points along the paramagnet phase boundary.  The errors on the exponents are estimated from fits performed at $T=T_N \pm \Delta T_N$.}
\vspace{-0.2cm}
\begin{tabular}{ccccccc}
\\
\hline\hline\
$H$&& $T_N$ && $\beta$ & $\gamma$ & $\nu$ \\
\hline\hline\
0&& 10.651(4) && 0.35(2)  & 1.33(5) & 0.670(7)       \\ 
0.10&& 10.651(5) && 0.31(3)  & 1.3(1) & 0.63(1)       \\
0.25&& 10.685(5) && 0.28(3)  & 1.22(6) & 0.56(1)        \\
0.50&& 10.756(5) && 0.27(4)  & 1.02(9) & 0.50(2)        \\
1.00&& 10.891(5) && 0.27(7)  & 1.0(1) & 0.48(4)     \\
\hline\hline
\end{tabular}
\end{center}
\vspace{-0.55cm}
\end{table}
%\begin{table*} 
%\vspace{-0.2cm}
%\begin{center}
%\caption{\label{critexp2} Critical temperature and critical exponents for several points along the paramagnet phase boundary.  The %errors represent the robustness of the fits. The set of exponents $\beta^\pm$, $\gamma^\pm$ and $\nu^\pm$ are obtained from the fits %performed at $T=T_N\pm\Delta T_N$. }
%\vspace{-0.2cm}
%\begin{tabular}{ccccccccccccccccccc}
%\\
%\hline\hline\
%$H$&&& $T_N$ &&& $\beta$ & $\gamma$ & $\nu$ &&& $\beta^-$ & $\gamma^-$ & $\nu^-$ &&& $\beta^+$ & $\gamma^+$ & $\nu^+$  \\
%\hline\hline\
%0&&& 10.651(4) &&& 0.355(6)  & 1.33(3) & 0.670(7) &&& 0.341(7)&1.38(3)&0.671(6)&&& 0.370(6)&1.28(2)&0.669(7)        \\ 
%0.10&&& 10.651(5) &&& 0.305(8)  & 1.31(3) & 0.63(1) &&& 0.29(1)&1.45(7)&0.64(1)&&& 0.33(1)&1.32(5)&0.64(1)        \\
%0.25&&& 10.685(5) &&& 0.279(9)  & 1.22(4) & 0.56(1) &&& 0.26(1)&1.28(4)&0.56(1)&&& 0.306(6)&1.17(2)&0.57(1)       \\
%0.50&&& 10.756(5) &&& 0.266(7)  & 1.02(2) & 0.50(1) &&& 0.23(1)&1.11(3)&0.48(1)&&& 0.306(6)&0.94(2) &0.52(1)       \\
%1.00&&& 10.891(5) &&& 0.269(7)  & 0.96(3) & 0.48(1) &&& 0.21(2)&1.10(4)&0.45(1)&&& 0.337(7)&0.90(3) &0.52(1)       \\
%\hline\hline
%\end{tabular}
%\end{center}
%\vspace{-0.7cm}
%\end{table*}

The effects of a nonzero applied field were examined here for the quasi-1D case II only.  This extends the work of Refs. \cite{Netz1,Plumer1} where isotropic exchange was assumed and where it is argued the effect of an applied field is to change the symmetry to that of the 3-state Potts model and hence the transition should be first order within mean field theory. However, a more complicated phase diagram emerged as a result of these earlier MC simulations. The same number of MC steps as in the zero field cases were used in the present study for equilibration and to calculate thermal averages. At field strengths $H=$ 0.01, 0.25, 0.50 and 1.0 simulation results for the energy cumulant extrapolate to $U_E^*=0.666663(3)$, suggesting that the phase transitions remain continuous at these lower values.   Results for the critical exponents at five field strengths from $H=0$ to $H=1$ are shown in Table \ref{critexp2}.     The error for the critical temperature is estimated from the scatter of the crossing points of the order-parameter cumulant data. The errors on the critical exponents are obtained from fits performed at $T=T_N\pm\Delta T_N$.   For example, for $H=0.25$, the fits performed at $T=10.680$ yielded $\beta=0.26(1)$, $\gamma=1.28(4)$, $\nu=0.56(1)$, whereas those at $T=10.690$ yielded $\beta=0.306(6)$, $\gamma=1.17(2)$, $\nu=0.57(1)$.    These results illustrate the sensitivity of the critical exponents with respect to the choice of critical temperature.  The largest source of error on the critical exponents comes from the uncertainty of the critical temperature.   The errors from the robustness of the fits are typically smaller.    For small values of the applied field ($H=0.1$) the estimated values of the critical exponents $\beta$ and $\nu$ are slightly lower than the zero field cases. At intermediate field values $H=0.5$ and $H=1.0$ the magnetic phase transition is characterized by a set of exponents that are consistent with tricritical mean-field values.  Monte Carlo simulations results for $H=2.0$ reveal a weak first order phase transition where $U_E^*=0.666642(3)$.
%\begin{figure}
%\begin{center}
%\includegraphics[width=0.45\textwidth]{c2.eps}\\
%\caption{\label{c2} Scaling of the energy cumulant $U_E$ as a function of lattice size $L$ is shown for the several values of the applied field $H$.  The horizontal line denotes $U_E=2/3$.  Need to add results for $H=2.0$. } 
%\end{center}
%\end{figure}
%%%%%%%%%%%%%%%%%%%%%%%%%%%%%%%%%%%%%%%%%%%%%%%%%%%%%%%%%%%%%%%%%%%%%%%%%%%%%%%%%%%%%%%%%%%%%%%%%%%%%%%%%%%%%%%%%%%
%%%%%%%%%%%%%%%%%%%%%%%%%%%%%%%%%%%%%%%%%%%%%%%%%%%%%%%%%%%%%%%%%%%%%%%%%%%%%%%%%%%%%%%%%%%%%%%%%%%%%%%%%%%%%%%%%%%

The results of these extensive MC simulations using the CHB algorithm combined with the histogram analysis technique serve to resolve long-standing questions regarding both experimental and previous modeling results on the critical properties of the quasi-1D ISTAF where both XY and tricritical behavior has been reported.  This work illustrates the difficulty in extracting reliable estimates of exponents due to the enhanced sensitivity of the critical region when exchange interactions are frustrated and anisotropic.  A major focus of the present work has been the careful estimation of errors due to uncertainty in the assumed critical temperature, a feature relevant for the analysis of both modeled and measured data. Our results make clear that the classical model system yields XY criticality in the case of zero applied field for a variety of assumed exchange interactions, in agreement with most previous MC simulations results and with most experimental results on Ising-like ABX$_3$ compounds such as $S=1$ CsNiCl$_3$ \cite{collins}. Quantum effects due to SRO inherent in the quasi-1D $S=1/2$ compounds are mimicked by the addition of a small applied field and are shown to induce a cross-over to tricritical, then first-order behavior.  Our results explain recent high-resolution neutron scattering experiments indicating tricritical exponents in CsCoBr$_3$ \cite{Mao1} where the strength of the extra field term had previously been estimated to be $H/J_0 =0.05$ \cite{Nagler1} consistent with our simulations results ($H=0.5$).  This conclusion illustrates that quantum spin effects can control the experimentally accessible effective critical behaviour of quasi-1D ISTAF's.  This result is relevant to other systems that exhibit strong SRO before the onset of LRO.

We thank B. Southern and S. Nagler for insightful discussions. This work was supported by the Natural Sciences and Engineering Research Council of Canada (NSERC) and the Atlantic Computational Excellence Network (ACEnet).

\end{document}